\documentclass[default,iicol]{sn-jnl}

\usepackage{graphicx}%
\usepackage{multirow}%
\usepackage{amsmath,amssymb,amsfonts}%
\usepackage{amsthm}%
\usepackage{mathrsfs}%
\usepackage[title]{appendix}%
\usepackage{xcolor}%
\usepackage{textcomp}%
\usepackage{manyfoot}%
\usepackage{booktabs}%
\usepackage{algorithm}%
\usepackage{algorithmicx}%
\usepackage{algpseudocode}%
\usepackage{listings}%
\usepackage{threeparttablex} 

\begin{document}

\title[Article Title]{Shape evolution in even-mass $^{98-104}$Zr isotopes via lifetime measurements using the $\gamma\gamma$-coincidence technique}


\author[1]{G.~Pasqualato} 
\author[2]{S.~Ansari} 
\author[3]{J.S.~Heines} 
\author[3]{V.~Modamio} 

\author[3]{A.~G\"orgen} 
\author[2]{W.~Korten} 
\author[1,4]{J.~Ljungvall} 

\author[5]{E.~Clément} 
\author[6]{J.~Dudouet} 
\author[5]{A.~Lemasson}
\author[7]{T.R.~Rodr\'iguez} 

\author[8]{J.M.~Allmond}
\author[9]{T.~Arici}
\author[3]{K.S.~Beckmann}
\author[10]{A.M.~Bruce}
\author[11]{D.~Doherty}
\author[12]{A.~Esmaylzadeh}
\author[10]{E.R.~Gamba}
\author[12]{L.~Gerhard}
\author[9]{J.~Gerl}
\author[1]{G.~Georgiev}
\author[13]{D.P.~Ivanova}
\author[12]{J.~Jolie}
\author[5]{Y.-H.~Kim}
\author[12]{L.~Knafla}
\author[1]{A.~Korichi}
\author[9]{P.~Koseoglou}
\author[14]{M.~Labiche}
\author[13]{S.~Lalkovski}
\author[15]{T.~Lauritsen}
\author[5]{H.-J.~Li}
\author[3]{L.G. Pedersen}  
\author[9]{S.~Pietri}
\author[1]{D.~Ralet}
\author[12]{J.M.~Regis}
\author[11]{M.~Rudigier}
\author[9]{S.~Saha}
\author[3]{E.~Sahin}
\author[3]{S.~Siem}
\author[2]{P.~Singh}
\author[9,16,17]{P.-A.~Söderström}
\author[2]{C.~Theisen}
\author[18]{T.~Tornyi}
\author[2]{M.~Vandebrouck}
\author[16]{W.~Witt}
\author[2]{M.~Zieli\'nska}

\author[19]{D.~Barrientos}
\author[20]{P.~Bednarczyk}
\author[21]{G.~Benzoni}
\author[22]{A.J.~Boston}
\author[22]{H.C.~Boston}
\author[21,23]{A.~Bracco}
\author[24]{B~ Cederwall}
\author[20]{M.~Ciemala}
\author[5]{G.~de~France}
\author[25]{C.~Domingo-Pardo}
\author[12]{J.~Eberth}
\author[25]{A.~Gadea}
\author[26]{V.~González}
\author[27]{A.~Gottardo}
\author[22]{L.J.~Harkness-Brennan}
\author[12]{H.~Hess}
\author[22]{D.S.~Judson}
\author[7]{A.~Jungclaus}
\author[28,29]{S.M.~Lenzi}
\author[21,23]{S.~Leoni}
\author[28]{R.~Menegazzo}
\author[28,29]{D.~Mengoni}
\author[5,30]{C.~Michelagnoli}
\author[27]{D.R.~Napoli}
\author[31]{J.~Nyberg}
\author[11]{Zs.~Podolyak}
\author[21,23]{A.~Pullia}
\author[28,29]{F.~Recchia}
\author[12]{P.~Reiter}
\author[4,28]{K.~Rezynkina}
\author[2]{M.D.~Salsac}
\author[26]{E.~Sanchis}
\author[32]{M.~Şenyiğit}
\author[2,15,27]{M.~Siciliano}
\author[13]{J.~Simpson}
\author[18]{D.~Sohler}
\author[6]{O.~Stezowski}
\author[27]{J.J.~Valiente-Dobón}
\author[1]{D.~Verney.}


\affil[1] {IJCLab, IN2P3/CNRS, Université Paris-Saclay, Orsay, France}
\affil[2] {IRFU, CEA, Université Paris-Saclay, Gif-sur-Yvette, France}
\affil[3] {Department of Physics, University of Oslo, Oslo, Norway}
\affil[4] {Université de Strasbourg, CNRS, IPHC UMR 7178, Strasbourg, France}
\affil[5] {Grand Accélérateur National d’Ions Lourds, CEA/DRF-CNRS/IN2P3, Caen, France}
\affil[6] {Université Lyon, Université Lyon-1, CNRS/IN2P3, UMR5822, IP2I, Villeurbanne, France}
\affil[7] {Departamento de Estructura de la Materia, Física Térmica y Electrónica and IPARCOS, Universidad Complutense de Madrid, Madrid, Spain}
\affil[8] {Physics Division, Oak Ridge National Laboratory, Oak Ridge, TN, USA}
\affil[9] {GSI Helmholtzzentrum für Schwerionenforschung GmbH, Darmstadt, Germany}
\affil[10] {School of Computing, Engineering and Mathematics, Brighton University, Brighton, UK}
\affil[11] {Department of Physics, University of Surrey, Guildford, UK}
\affil[12] {Institut für Kernphysik, Universität zu Köln, Köln, Germany}
\affil[13] {Faculty of Physics, University of Sofia, Sofia, Bulgaria}
\affil[14] {STFC Daresbury Laboratory, Daresbury, Warrington, UK}
\affil[15] {Argonne National Laboratory, Argonne, IL, USA}
\affil[16] {Department of Physics, Institute for Nuclear Physics, TU Darmstadt, Darmstadt, Germany}
\affil[17] {ELI-NP/IFIN-HH, Bucharest-Măgurele, Romania}
\affil[18] {Institute for Nuclear Physics, University of Debrecen, Debrecen, Hungary}
\affil[19] {CERN, CH-1211 Geneva 23, Switzerland}
\affil[20] {The Henryk Niewodniczanski Institute of Nuclear Physics, Kraków, Poland}
\affil[21] {INFN Sezione di Milano, Milano, Italy}
\affil[22] {Oliver Lodge Laboratory, The University of Liverpool, Liverpool, UK}
\affil[23] {Dipartimento di Fisica, Università di Milano, Milano, Italy}
\affil[24] {Department of Physics, KTH Royal Institute of Technology, Stockholm, Sweden}
\affil[25] {Instituto de Física Corpuscular, CSIC-Universidad de Valencia, Valencia, Spain}
\affil[26] {Departamento de Ingeniería Electrónica, Universitat de Valencia, Burjassot, Spain}
\affil[27] {INFN Laboratori Nazionali di Legnaro, Legnaro, Padova, Italy}
\affil[28] {INFN Sezione di Padova, Padova, Italy}
\affil[29] {Dipartimento di Fisica e Astronomia dell’Università di Padova, Padua, Italy}
\affil[30] {Institut Laue-Langevin, Grenoble, France}
\affil[31] {Department of Physics and Astronomy, Uppsala University, Uppsala, Sweden}
\affil[32] {Department of Physics, Ankara University, Besevler, Ankara, Turkey}


\abstract{The Zirconium (Z=40) isotopic chain has attracted interest for more than four decades. 
The abrupt lowering of the energy of the first $2^+$ state and the increase in the transition strength B(E2;~$2^+_1\rightarrow0^+_1)$ going from $^{98}$Zr to $^{100}$Zr has been the first example of ``quantum phase transition'' in nuclear shapes, which has few equivalents in the nuclear chart. 
Although a multitude of experiments have been performed to measure nuclear properties related to nuclear shapes and collectivity in the region, none of the measured lifetimes were obtained using the Recoil Distance Doppler Shift method in the $\gamma\gamma$-coincidence mode where a gate on the direct feeding transition of the state of interest allows a strict control of systematical errors. 
This work reports the results of lifetime measurements for the first yrast excited states in $^{98-104}$Zr carried out to extract reduced transition probabilities. 
The new lifetime values in $\gamma\gamma$-coincidence and $\gamma$-single mode are compared with the results of former experiments.
Recent predictions of the Interacting Boson Model with Configuration Mixing, the Symmetry Conserving Configuration Mixing model based on the Hartree-Fock-Bogoliubov approach and the Monte Carlo Shell Model are presented and compared with the experimental data.
}




\maketitle

\section{Introduction}\label{introduction}

\begin{figure}[bp]
\centerline{%
\includegraphics[width=0.5\textwidth]{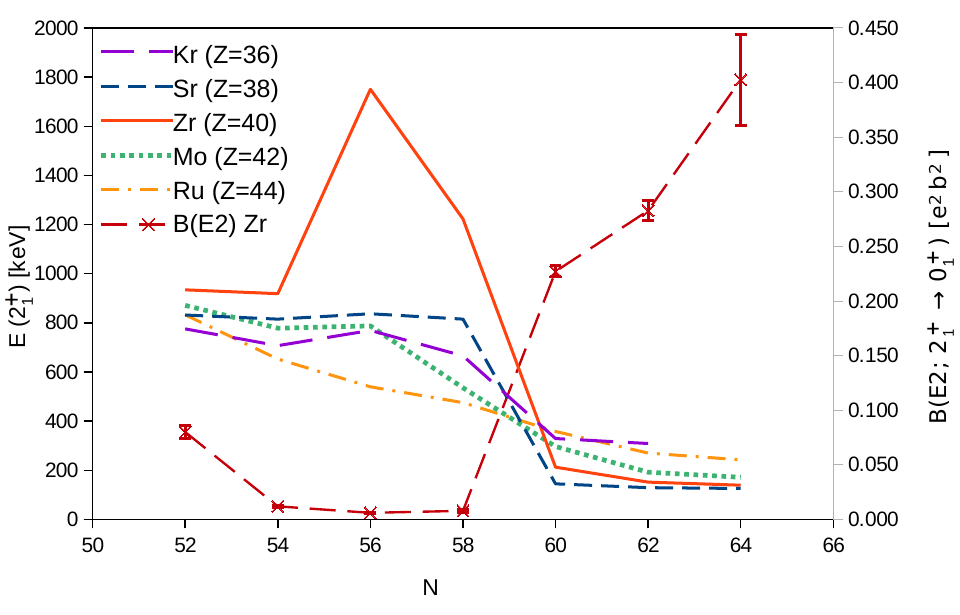}}
\caption{Evolution of the $2^+_1$ excitation energy as a function of neutron number for even-even nuclei in the A$\sim$100 region. The experimental B(E2;~$2^+_1 \rightarrow 0^+_1$) for Zr isotopes are also reported. The values are taken from the Evaluated Nuclear Structure Data File~\cite{nndc}.}
\label{fig:energy_BE2}
\end{figure}

The neutron-rich nuclides around A$\sim$100 represent a well-established example of static nuclear deformation, as predicted in 1969 by the theoretical work of D.A.~Arseniev \textit{et al}.~\cite{Arseniev1969}.
The first experimental evidence of stable deformation in the mass range 92-110 was found by S.~A.~E. Johansson in 1965~\cite{Johansson1965} within the study of the $\gamma$ radiation emitted from fission fragments of $^{252}$Cf. 
It was confirmed shortly after by E.~Cheifetz \textit{et al.}~\cite{Cheifetz1970} with the  rotational-like behaviour in even-even Zr, Mo, Ru and Pd nuclei as given from the systematic of the energy spacing between excited states. 
The comparison with the results from a contemporary $^{96}$Zr(t,p)$^{98}$Zr experiment~\cite{Blair1969}, revealed a surprising change in the energy of the lowest $2^+$ states between $^{98}$Zr and $^{100}$Zr, as depicted in
\figurename~\ref{fig:energy_BE2}.
These measurements reveal that the collectivity, which tends to evolve in a gradual way throughout the nuclear chart, increases drastically for Zr isotopes heavier than $^{98}$Zr. 
A similar, but less drastic, trend is also found for the neighbouring even-even Sr (Z=38) isotopes. 
However, the shape transition at N=60 becomes less abrupt as the distance to the proton harmonic oscillator shell Z=40 increases, which can be observed already for Mo, Kr and Ru nuclei (see \figurename~\ref{fig:energy_BE2}).
In addition to the energy of the first $2^+$ states, other nuclear properties such as electromagnetic moments, two-neutron separation energies and the energy ratios E$(4_1^+)$/E$(2_1^+)$ can be used as measures of the collectivity. As an example, the B(E2;~$2^+_1 \rightarrow 0^+_1$) transition strengths for Zr isotopes, presented in \figurename~\ref{fig:energy_BE2}, show a sudden increase after N=58.

Many theoretical studies attempted to reproduce and understand the rapid increase of collectivity appearing at N=60 in the zirconium and strontium nuclei. 
According to shell model studies by Federman, Pittel and collaborators~\cite{FedermanPittel1977,Federman1979,FedermanPittel1979,Etchegoyen1989}, the increasing deformation is due to the strong residual interaction between protons and neutrons occupying the spin-orbit partner orbitals $0\nu g_{7/2}$ and $0\pi g_{9/2}$, which have a large spatial overlap. 
As a consequence, the effective single-particle energies of valence orbitals shift and promote multiple particle-hole (\hbox{\textit{p-h}}) excitations which act coherently to induce deformation. 
These results agree with self-consistent Hartree-Fock studies~\cite{Goodman1976,Goodman1977,Zeldes1983} which also underline the importance of the residual proton-neutron interactions driving major modifications in the occupation of the $0\nu g_{7/2}$ and $0\pi g_{9/2}$ orbitals.

While nuclear theory agrees in the description of the onset of deformation in Zr isotopes with increasing neutron number, the sharp change in the spectroscopic properties between $^{98}$Zr and $^{100}$Zr has been extremely difficult to reproduce. 
This is particularly true for the E2 transition strengths between low-lying excited states. 
For a detailed review of the theoretical framework concerning Zr isotopes, the reader is referred to Ref.~\cite{GarciaRamos2019} and for a recent experimental overview to Ref.~\cite{Garrett2022}. 
Mean field studies, using the self-consistent Hartree-Fock-Bogoliubov (HFB) approach, predict a smooth evolution of the transition strength with increasing neutron number with enhanced B(E2;~$2^+_1 \to 0^+_1$) already observed around N=54 ~\cite{Delaroche2010,Mei2012}. 
Large scale shell model (LSSM) calculations~\cite{Sieja2009} are able to correctly predict the small B(E2;~$2^+_1 \rightarrow 0^+_1$) values up to $^{98}$Zr. 
To address the increase in collectivity the use of a very large valence space is required and heavier Zr isotopes are presently out of reach for these calculations.

Monte Carlo Shell Model calculations (MCSM)~\cite{Honma1995,Otsuka1998} have proven capable in reproducing nuclear spectroscopic data with good precision.
In the MCSM calculations of Togashi \textit{et al.}~\cite{Togashi2016}, a multitude of distinct structures coexisting at low excitation energy was predicted in zirconium isotopes around N=60. 
In particular, the ground states up to $^{98}$Zr are calculated to have a spherical configuration, with few protons in the $0g_{9/2}$ orbital. 
A prolate-deformed structure, characterized by a large occupancy of the $0\pi g_{9/2}$ orbital, appears as the first excited $0^+$ state in $^{98}$Zr and becomes the ground state from $^{100}$Zr onward.  
Deformed structures arise in particular as a consequence of the lowering of the $0\nu g_{7/2}$ and $0\nu h_{11/2}$ orbitals, enabling the development of quadrupole correlations.
As explained in the work of Togashi \textit{et al.}~\cite{Togashi2016}, quadrupole correlations are mainly due to the tensor part of the nucleon-nucleon interaction~\cite{Otsuka2016}. 
This large difference in configuration between the two structures is the key to the abrupt change, since any mixing between them would have smoothed out the nuclear properties.

Observation of excited $0^+$ states at low excitation energy hints at possible shape coexistence. Such states were identified in $^{98,100}$Zr~\cite{Kawade1982,Wohn1986}, as well as in $^{96,98}$Sr~\cite{Schussler1980,Jung1980} and in the latter, shape coexistence and configuration inversion at N=60 has been experimentally confirmed via detailed Coulomb-excitation studies~\cite{Clement2016_PRL,Clement2016_PRC}.
The recent work of P.~Singh \emph{et al.}~\cite{Singh2018} measured the lifetime of the $2_1^+$, $4_1^+$ and $6_1^+$ states in $^{98}$Zr and the coexistence of three distinct shapes (a spherical $0^+_1$ state, a prolate deformed $0_2^+$ state and a triaxial $0_3^+$ state) was proposed in conjunction with MCSM calculations.

The rapid evolution of nuclear properties in this region has been described as a nuclear quantum phase transition (QPT)~\cite{GilmoreFeng1978} following the similarities with thermodynamic phase transitions. 
Quantum phase transitions of nuclear shapes, in particular at the critical points, are reviewed in Ref.~\cite{Cejnar2010}.
The shape transition in the Zr isotopes is also described by the Interacting Boson Model~\cite{IachelloArima1987} in the configuration-mixing framework (IBM-CM), an extension of the IBM formulated by F.~Iachello and A.~Arima. 
The configuration-mixing approach~\cite{DuvalBruce1981,DuvalBruce1982} treats simultaneously several boson configurations corresponding to different \textit{p-h} excitations across the shell closure. 
Configuration-mixed QPT and phenomena of shape coexistence in nuclei have been studied extensively in the IBM-CM framework~\cite{DuvalBruce1981,DuvalBruce1982,Frank2006,Sambataro1982,Nomura2016,DuvalBruce1983,PadillaRoda2003,Fossion2003,GarciaRamos2011,GarciaRamosHeyde2014,GarciaRamos2014,GarciaRamosHeyde2015,Leviatan2018} 
and recent IBM-CM calculations for Zr isotopes~\cite{GarciaRamos2020,Gavrielov2019,Gavrielov2022} successfully describe the trend of the B(E2;~$2^+_1 \rightarrow 0^+_1$) values by considering an intertwined quantum phase transition, involving the crossing of two configurations, each of which undergoes its own QPT. \\

To provide more stringent tests to the nuclear models describing the rapid changes of collectivity in the Zr isotopes around N=60, experimental data on the electromagnetic properties of low-lying excited nuclear states are of key importance.
These results will allow us to understand how the deformation or the rotational behaviour evolves with spin.
It is therefore important to measure them with high precision and small systematic uncertainties. 
Thanks to the increasing performance of $\gamma$-ray detector arrays, it has become possible to extract reduced transition probabilities in heavier Zr nuclei by using lifetime measurements in $\gamma\gamma$-coincidence mode in order to strongly limit the systematic errors which affect lifetime measurements in $\gamma$-single mode. 
The present manuscript reports the results of recent lifetime measurements with the Recoil Distance Doppler Shift (RDDS) technique~\cite{Dewald2012} in $^{98}$Zr, $^{100}$Zr, $^{102}$Zr and $^{104}$Zr. 
The measurement of the lifetimes with the $\gamma\gamma$-coincidence technique is performed for the first time in this work for nuclei in this mass region and results are compared with previous measurements and theoretical calculations.

\section{Experimental procedure} \label{exp_procedure}

To study electromagnetic properties of the neutron-rich nuclei around A$\sim$100, an experiment was performed at GANIL by using the $\gamma$-ray tracking-array AGATA~\cite{Akkoyun2012,Clement2017} and the magnetic spectrometer VAMOS++~\cite{Rejmund2011}. 
A $^{238}$U beam was accelerated by the separated sector cyclotron CSS1~\cite{Leherissier2004} to 6.2 MeV/u and directed onto a target of $^9$Be of 1.85~mg/cm$^2$ thickness inducing a fusion-fission reaction. 
AGATA is an array of position-sensitive high-purity Ge detectors, each of which is 36-fold segmented. 
In this experiment AGATA was composed of 1 double and 13 triple modules forming a total of 41 HPGe crystals, positioned at backward angles (from $\sim 135^\circ$ to $175^\circ$ with respect to the VAMOS++ axis) in order to detect de-excitation $\gamma$ rays with a maximized Doppler shift. 
The digitized pulse shapes coming from the segmented HPGe crystals are compared with a data base of simulated detector responses using the Pulse Shape Analysis (PSA) technique~\cite{Venturelli2005}.
This technique enables the accurate determination of all
$\gamma$-ray interaction points inside a HPGe crystal. 
The reconstruction of Compton-scattered $\gamma$ rays inside the array is then performed thanks to a tracking algorithm~\cite{LopezMartens2004}. 
For a complete description of the data analyses procedures see, e.g., Ljungvall \textit{et al.}~\cite{Ljungvall2020}.
The detection of $\gamma$ rays in coincidence with the fission fragment of interest enables the analysis of $\gamma$-ray spectra and $\gamma\gamma$ matrices of a single nuclide. 
For this purpose, the VAMOS++ spectrometer was used in dispersive mode: a magnetic field, created by the dipole, separates the particles along the horizontal axis according to their momentum and charge state. 
The position of VAMOS++ at 19$^\circ$ with respect to the beam direction ensures maximum detection efficiency for the fission fragments around A$\sim$100 (B$\rho = 1.11$~Tm). 
The trajectory of a particle inside VAMOS++ is reconstructed thanks to a Dual Position-sensitive Multi-Wire Proportional Counter (DP-MWPC) at the entrance of the spectrometer and two Drift Chambers (DC) at the focal plane. 
Here, a Multi-Wire Parallel-Plate Avalanche Counter (MWPPAC) together with the DP-MWPC measure the time of flight. Downstream of the MWPPAC, a set of six consecutive ionization chambers (IC) measure the total energy and the characteristic energy loss of the fragments.
Further details on the experiment and the analysis techniques can be found in Refs.~\cite{Rejmund2011,Hagen2017,Kim2017}.
All the information required to identify the reaction products in terms of mass and atomic number are acquired on an event-by-event basis. 

The Orsay Universal Plunger system~\cite{Ljungvall2012} was installed in the reaction chamber hosting the target and a $^{nat}$Mg degrader, which could be placed at an adjustable distance from the target allowing RDDS measurements for lifetimes from a few ps to hundreds of ps.
The RDDS technique~\cite{Dewald2012} is based on the separation, for the same $\gamma$-ray transition, of two components with a different Doppler-shifted energy, corresponding to the $\gamma$ emission of recoiling nuclei between the target and the degrader, with a velocity $v_{in}$, and after the degrader, with a different velocity $v_{out}$.
The thickness of the degrader, 4.5~mg/cm$^2$, ensured a sufficient separation between the two Doppler-shifted components thanks to the change of velocity of the emitting nucleus in the degrader and, at the same time, did not affect substantially the shapes of the peaks in the $\gamma$-ray spectra because of the slowing-down process. 
Likewise, the kinetic energy loss of the recoiling nuclei in the degrader was limited, enabling proper identification of the fragments in VAMOS++. 
Data was collected for ten plunger distances ranging from 30 to 2650 $\mu$m.

\section{Analysis} \label{analysis}
\begin{figure}[htbp]
\centerline{%
\includegraphics[width=0.46\textwidth]{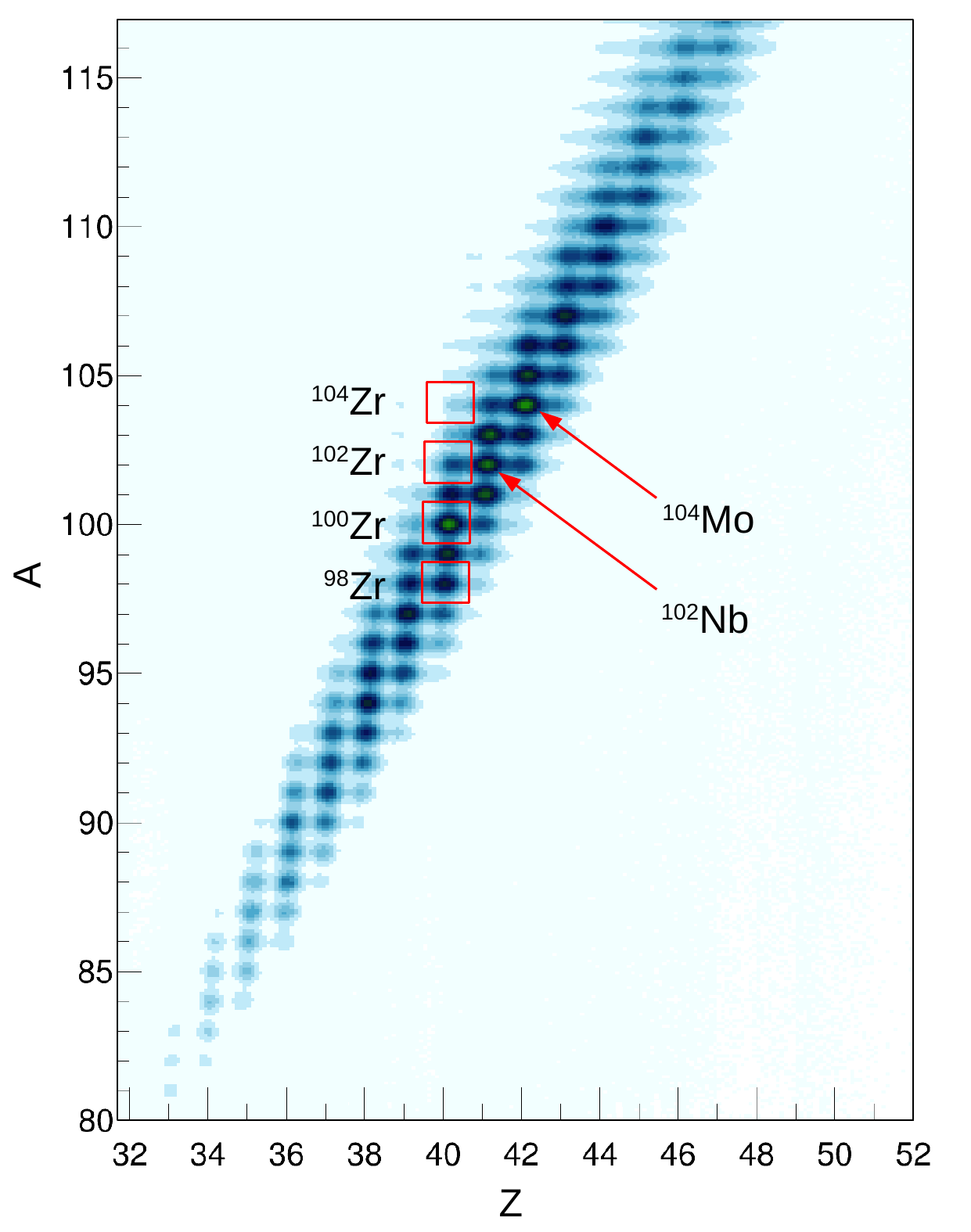}}
\caption{Correlation matrix between the mass A, summed over all charge states, and the atomic number Z measured by VAMOS++ over the full focal plane. The most populated fission fragments ($^{104}$Mo, $^{102}$Nb) and the four analyzed systems ($^{98}$Zr, $^{100}$Zr, $^{102}$Zr, $^{104}$Zr) are indicated for reference. }
\label{fig:VAMOS_MZ}
\end{figure}

\begin{figure}[htbp]
\centerline{%
\includegraphics[width=0.48\textwidth]{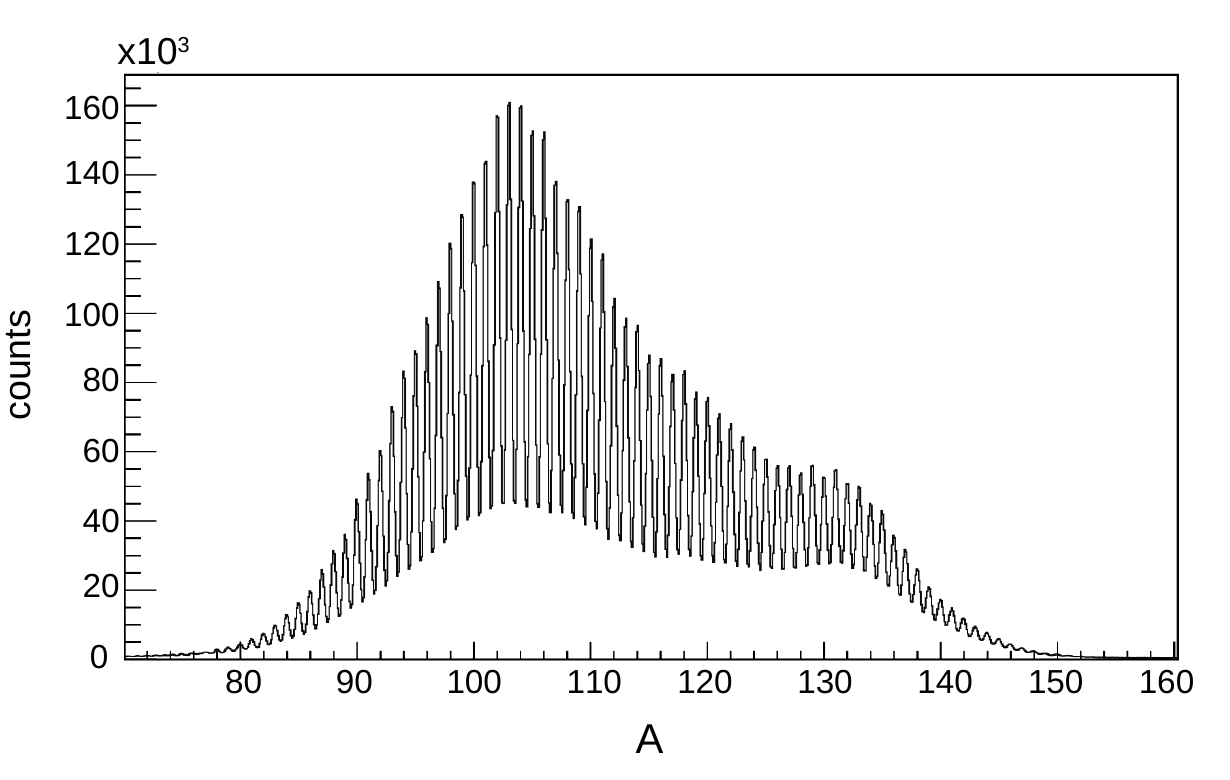}}
\caption{Mass spectrum summed over all the charge states from the VAMOS++ identification.}
\label{fig:VAMOS_mass}
\end{figure}

The $\gamma$ rays belonging to the recoiling nucleus of interest can be selected by gating in the correlation matrix between the mass A and the atomic number Z obtained from the VAMOS++ identification.
\figurename~\ref{fig:VAMOS_MZ} presents the A $vs$ Z matrix and shows that the large acceptance in the magnetic rigidity $B\rho$ ($\pm 10\%$) of VAMOS++ enables the identification of many of the produced isotopes with significant statistics. 
The most populated isotopes are $^{100}$Zr, $^{102}$Nb and $^{104}$Mo. \figurename~\ref{fig:VAMOS_mass} presents a one-dimensional mass spectrum, showing that nuclei with mass numbers between A$\sim 100-108$ are the strongest channels.

The fission fragments produced with a $^{238}$U beam in inverse kinematics have a velocity ($v_{out}$) of about 10\% of the speed of light, which assures significant Doppler shifts on the emitted $\gamma$ rays. 
The direction and the velocity of the recoiling nuclei after the degrader is determined from the reconstructed trajectory and the time-of-flight information measured with VAMOS++.
This velocity is used to correct the energy of $\gamma$ rays for the Doppler effect on an event-by-event basis. In this way, the energy of $\gamma$ rays emitted after the degrader have the nominal energy of the $\gamma$-ray transition (unshifted component). 
Whereas, if the nucleus decays before the degrader, the energy is shifted from the nominal value (shifted component) due to a different velocity of the recoil. 
The fission fragments which enter the VAMOS++ spectrometer have a velocity distribution, centred around the mean velocity $\langle v \rangle$, which depends on the beam energy, the reaction mechanism and the target and degrader thicknesses.
The observed velocity distribution $v_{out}$ depends on the angular and the momentum acceptance of VAMOS++. 
For the analysis of RDDS data the velocity of the recoil before the degrader is required: the velocity distribution $v_{in}$ of the recoils before the degrader is obtained by adding a constant velocity shift $\delta v = v_{in} - v_{out}$ which results from the mean energy loss in the degrader.
The $\delta v$ factor is determined directly from the Doppler shift of different known transitions by measuring the energy separation of the shifted and unshifted components in the $\gamma$-ray spectra.
The velocity difference $\delta v$ is about 15\%, sufficient for a complete separation of the unshifted and shifted components in the $\gamma$-ray spectra for the energy range of interest ($\sim 300 - 1300$~keV). 
For $^{98-104}$Zr the measured velocities are listed in \tablename~\ref{tab:velocities}.
\begin{table}[htbp]
    \caption{\label{tab:velocities} Measured velocities before ($v_{in}$) and after ($v_{out}$) the degrader for $^{98-104}$Zr. The values in the brackets correspond to the standard deviation of the velocity distribution.}
\begin{tabular}{p{2.cm}p{2.cm}p{2.cm}}
        \hline \hline
        \noalign{\smallskip}isotope & $v_{in}$ [$\mu$m/ps] & $v_{out}$ [$\mu$m/ps] \\
           \noalign{\smallskip} \hline
        
        \noalign{\smallskip}$^{98}$Zr  & 37.4(23) & 31.8(20) \\
        $^{100}$Zr & 37.0(23) & 31.4(21)   \\
        $^{102}$Zr & 36.3(24) & 30.8(22)  \\
        $^{104}$Zr & 35.2(25) & 29.6(23) \\
        \hline \hline
\end{tabular}
\end{table}

\begin{figure*}[htbp]
\centerline{%
\includegraphics[width=1.04\textwidth]{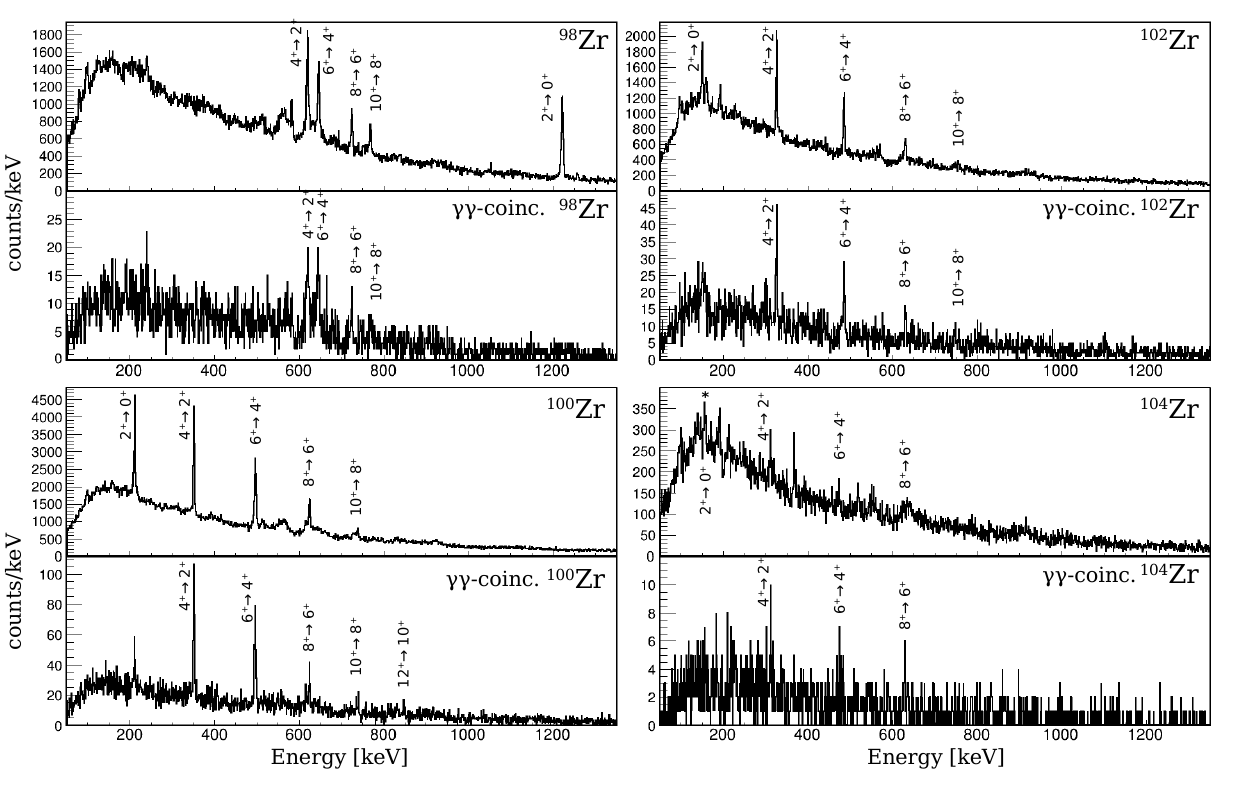}}
\caption{Doppler-corrected $\gamma$-ray spectra for $^{98}$Zr and $^{100}$Zr, on the left side, $^{102}$Zr and $^{104}$Zr, on the right side. For each nucleus $\gamma$-single  and $\gamma\gamma$-coincidence spectra (gated on the $2_1^+ \rightarrow 0_1^+$ transition) are reported at the shortest target-degrader distance (30~$\mu$m). The spectra are Doppler corrected using the velocity $v_{out}$ as in \tablename~\ref{tab:velocities}.}
\label{fig:all_spectra}
\end{figure*}

\begin{figure*}[htbp]
\centerline{%
\includegraphics[width=0.9\textwidth]{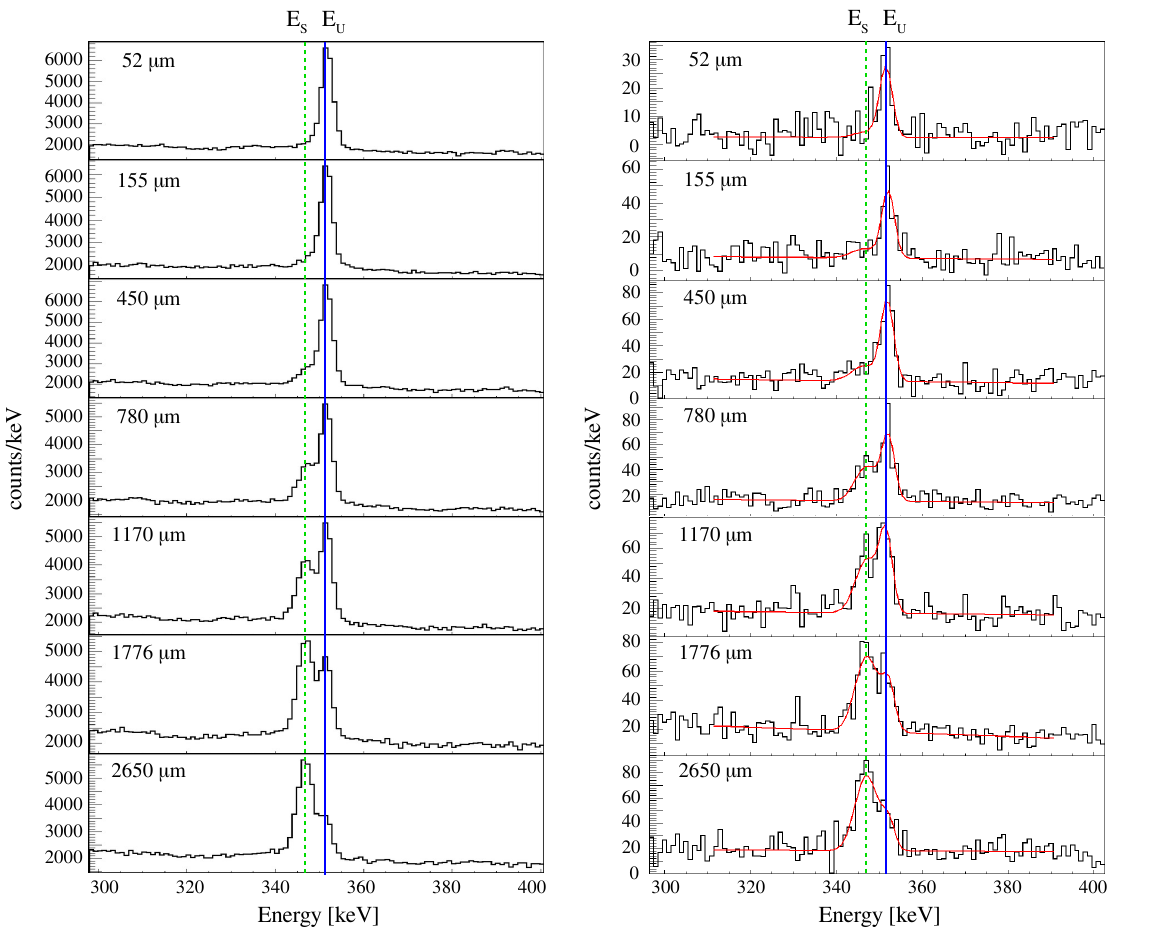}}
\caption{Gamma-ray spectra for seven target-degrader distances showing the $4_1^+ \rightarrow 2_1^+$ transition in $^{100}$Zr in $\gamma$-single mode (left-side panel) and in $\gamma\gamma$-coincidence mode (right-side panel) after selecting a 8-keV gate in the shifted component of the $6_1^+ \rightarrow 4_1^+$ direct feeding transition. The spectra are Doppler corrected using the velocity $v_{out}$ as in \tablename~\ref{tab:velocities}. The continuous (blue) and dashed (green) lines indicate the positions of the unshifted E$_U$ and shifted E$_S$ components, respectively. The continuous red lines in the $\gamma$-gated spectra show the functions used to fit the peaks as an example (see text for details).}
\label{fig:spectra100Zr}
\end{figure*}

The energy of the emitted $\gamma$ rays is recorded in $\gamma\gamma$ matrices obtained by selecting a certain area in the identification matrix (\figurename~\ref{fig:VAMOS_MZ}), corresponding to the fission fragment of interest. 
Examples of the Doppler-corrected $\gamma$-ray spectra are presented in \figurename~\ref{fig:all_spectra} for $^{98}$Zr, $^{100}$Zr, $^{102}$Zr and $^{104}$Zr both in $\gamma$-single and $\gamma\gamma$-coincidence mode. 
In this figure one can notice the excellent discrimination of the $\gamma$ rays from the fragment of interest made by VAMOS++.
The $\gamma$-single spectra can be compared with those in Ref.~\cite{Singh2018}.  
Employing the AGATA spectrometer instead of EXOGAM significantly increased the statistics, enabling the measurement of lifetimes from $\gamma\gamma$ coincidences, thus avoiding uncertainties in the results due to unobserved feeding transitions.
\figurename~\ref{fig:spectra100Zr} presents a part of Doppler-corrected $\gamma$-ray spectra of $^{100}$Zr for several target-degrader distances, in $\gamma$-single and $\gamma\gamma$-coincidence mode, showing the evolution of the ratio between shifted and unshifted components for the transition $4^+_1 \rightarrow 2^+_1$. 

From these data, the lifetimes of 15 excited states in the even-mass isotopes $^{98-104}$Zr are obtained using the Differential Decay Curve Method (DDCM)~\cite{Dewald2012}, of which 9 are measured in $\gamma\gamma$ coincidences with gates on transitions directly feeding the state of interest. 
The lifetime of the $6^+_1$ state in $^{102}$Zr and the $4^+_1$, $6^+_1$ states in $^{104}$Zr are measured for the first time. 
The analysis of $\gamma\gamma$ coincidences is advantageous when measuring the lifetime of low-lying excited states and the nucleus is populated at higher excitation energy. 
The coincidence with the in-flight component of a direct feeding transition of the state of interest avoids the complication related to the analysis of all observed feeding transitions and the estimation of the contribution from unseen feedings. 

\begin{figure*}[htbp]
\centerline{%
\includegraphics[width=1.03\textwidth]{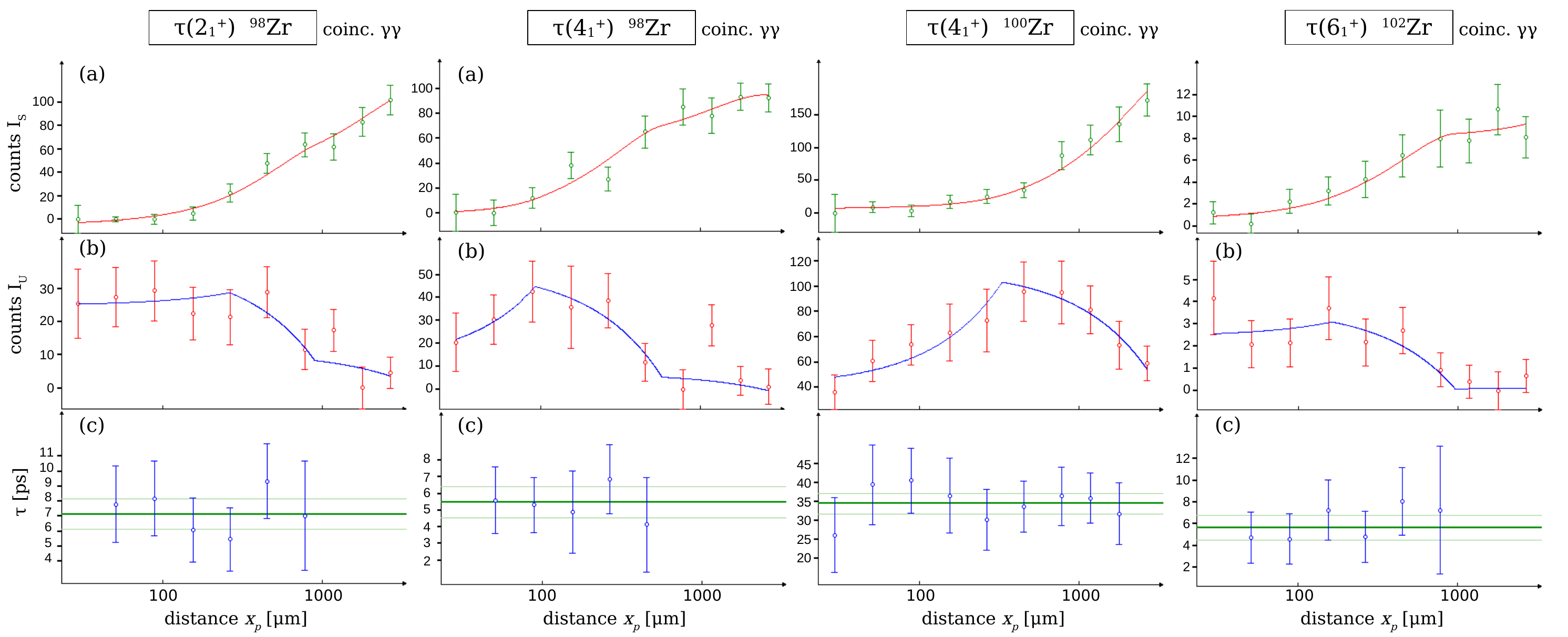}}
\caption{DDCM analysis for the measurement of the lifetime of the $2^+_1$ and the $4^+_1$ excited states in $^{98}$Zr, the $4^+_1$ in $^{100}$Zr and the $6^+_1$ in $^{102}$Zr in $\gamma\gamma$ coincidences with a gate on the shifted component of the direct feeding transition. Each panel reports: (a) normalized intensities of the shifted components I$_S$ with the fitting function $f(x)$ corresponding to three smoothly connected polynomials of second order; (b) normalized intensities of the stopped components I$_U$ with the fitting function proportional to the derivative of $f(x)$; (c) weighted average (thick green line) of the lifetimes $\tau(x_p)$ calculated at each plunger distance in the region of sensitivity, with the associated uncertainties shown with thin green lines.}
\label{fig:tau}
\end{figure*}

For each transition $A$, the intensities of the shifted $A_S$ and the unshifted $A_U$ components are measured in the projected spectrum, resulting from gating on the shifted component $F_S$ of a transition $F$ feeding the state of interest. 
The coincidence intensities $\{F_S,A_S\}$ and $\{F_S,A_U\}$ are measured from the projected spectra by using a gaussian fit where the parameters sigma and centroid are constrained to vary in an limited range. 
The lifetime of an excited state $\tau(x_p)$ at each plunger distance $x_p$ is obtained by applying the DDCM formula for the $\gamma\gamma$ analysis in coincidence with the direct feeding transition~\cite{Dewald2012}:
\begin{equation} \label{eq:tau}
    \tau(x_p)= \frac{ \{F_S,A_U\}(x_p)} {v_{in}\,\frac{d}{dx} \{F_S,A_S\}(x_p) } \, .
\end{equation} 
The intensities $\{F_S,A_U\}(x_p)$ and $\{F_S,A_S\}(x_p)$ in Eq.~(\ref{eq:tau}) are normalized in order to account for the different statistics collected at each plunger distance during the experiment. 
The normalization factor must be proportional to the number of reactions of interest in the target at each distance. 
Since the detection efficiency of VAMOS++ and AGATA is constant at each distance, we used as normalization factor the number of events in the projected $\gamma$-ray spectra after the gate in the VAMOS++ A $vs$ Z matrix.
The intensities of the shifted components $\{F_S,A_S\}(x_p)$ are fitted with a piece-wise polynomial function by using the software Napatau~\cite{napatau}. 
Following Eq.~\ref{eq:tau}, the derivative values of this curve, multiplied by the velocity of the recoil $v_{in}$ and the correct lifetime value $\tau(x_p)$, are equal to the intensity of the stopped components $\{F_S,A_U\}(x_p)$, for each plunger distance $(x_p)$.
The final value of the lifetime is the weighted average of the $\tau(x_p)$ values in the sensitive region of the technique. 
The region of sensitivity includes the distances where the shifted intensities  $I_s$ are strongly changing, which correspond indicatively to those plunger distances where the $\{F_S,A_U\}(x_p)$ curve is rising.
Examples of the DDCM analysis for the lifetime measurement in $\gamma\gamma$-coincidence mode for the $2^+_1$ and the $4^+_1$ excited states in $^{98}$Zr, the $4^+_1$ in $^{100}$Zr and the $6^+_1$ in $^{102}$Zr are presented in \figurename~\ref{fig:tau}. 

\begin{figure*}[htbp]
\centerline{%
\includegraphics[width=1.03\textwidth]{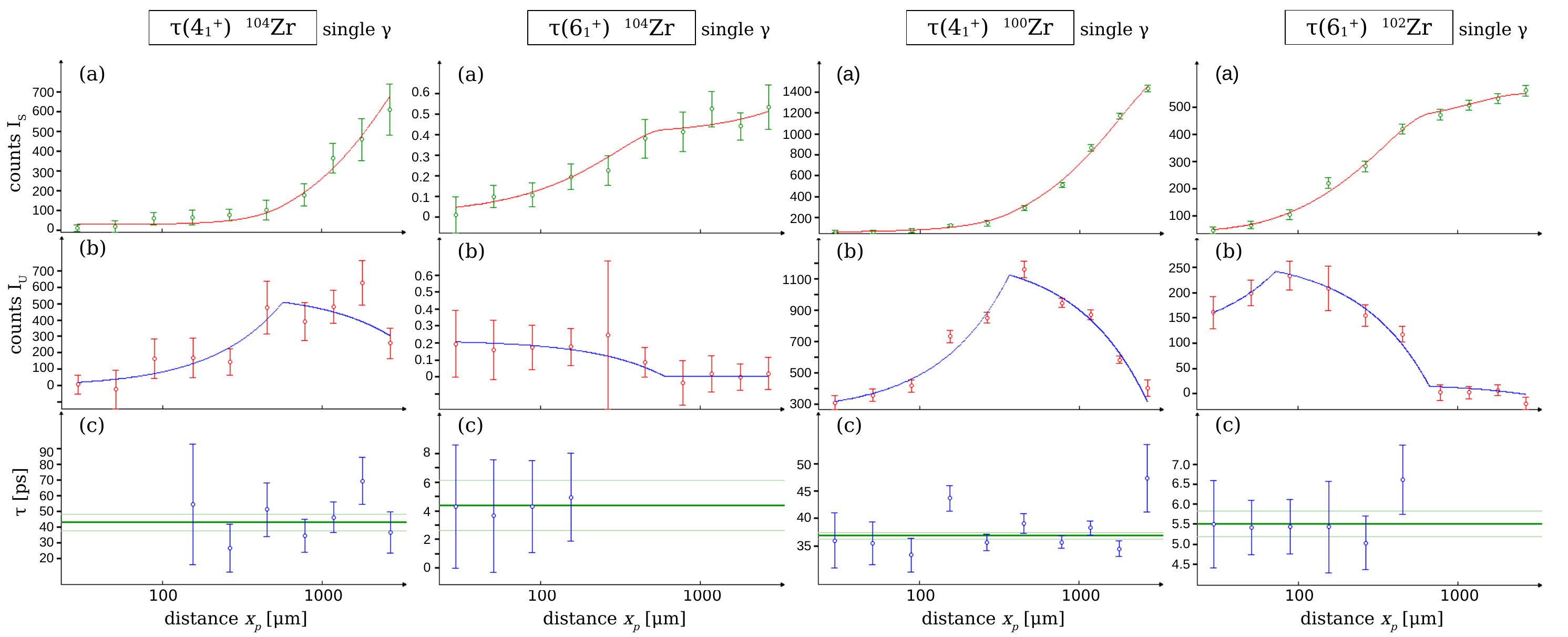}}
\caption{DDCM analysis for the measurement of the lifetime of the $4^+_1$ and the $6^+_1$ excited states in $^{104}$Zr, the $4^+_1$ in $^{100}$Zr and the $6^+_1$ in $^{102}$Zr in $\gamma$-single spectra. See \figurename~\ref{fig:tau} for the description of each panel.}
\label{fig:tau_single}
\end{figure*}

The lower statistics of higher-lying, less populated states prevents the measurement in $\gamma\gamma$ coincidences and a $\gamma$-single analysis was performed for these states instead. 
For the DDCM analysis in $\gamma$-single mode the intensities of all observed direct feeders $F_j$ have to be considered, multiplied by the proportionality factor $\alpha_j$ (introduced below) and the branching ratio $b_j$. 
In this approach, the normalization is included in the DDCM formalism~\cite{Dewald2012} as the sum of the shifted $I^S$ and unshifted $I^U$ intensities for each involved transition. In this case the ratio $Q=I^U/(I^U+I^S)$, i.e. the normalized unshifted intensities, is used to calculate the lifetime as follows:
\begin{equation} \label{eq:tau_single}
    \tau(x_p)= \frac{ -Q_A(x_p) + \sum_j b_j \alpha_j Q_{Fj}(x_p)} {v_{in}\,\frac{d}{dx} Q_A(x_p) } \, .
\end{equation}
The proportionality factor $\alpha_j$ is the weighted average of the quantities
\begin{equation} \label{eq:alpha}
    \alpha_j(x_p)= \frac{F_j^U(x_p)+F_j^S(x_p)}{A^U(x_p)+A^S(x_p)} \cdot \frac{\epsilon_A}{\epsilon_F} 
\end{equation} obtained at each plunger distance $x_p$, where $\epsilon_{A}$, $\epsilon_{F}$ are the detection efficiencies at the energy of the transition $A$ and $F_j$, respectively. 
As demonstrated in Ref.~\cite{Dewald2012}, Eq.~(\ref{eq:tau}) remains exact even in the presence of the deorientation effect. 
Although this effect is not eliminated when using DDCM in $\gamma$-single mode, it is significantly reduced in case the ratio $Q$ of the intensities is considered in Eq.~(\ref{eq:tau_single})~\cite{Dewald2012}. 
The consistency of the presented lifetime results in both $\gamma\gamma$-coincidence and $\gamma$-single mode confirms that the deorientation effect does not impact our data.

\figurename~\ref{fig:tau_single} shows the examples of the DDCM analysis for the lifetime measurement in $\gamma$-single mode for the $4^+_1$ and the $6^+_1$ excited states in $^{104}$Zr, the $4^+_1$ state in $^{100}$Zr and the $6^+_1$ state in $^{102}$Zr.

\begin{table*}[htbp]
\begin{center}
    \caption{\label{tab:results_lifetime}Mean-life results from the present experiment for excited states in $^{98}$Zr, $^{100}$Zr, $^{102}$Zr and $^{104}$Zr measured with the $\gamma\gamma$-coincidences and $\gamma$-single mode. Previously reported lifetime values $\tau_{lit.}$ for excited states $2^+$, $4^+$, $6^+$, $8^+$ and $10^+$ are listed for comparison. Energies E$_\gamma$ of the de-exciting transitions used for the measurement of the lifetime are also given. The energy values are taken from~\cite{nndc}.}
    \vspace{1mm}
\begin{tabular}{p{.9cm}p{1.9cm}p{1.9cm}p{2.1cm}p{7.cm}}
        \hline \hline
        \noalign{\smallskip}\multicolumn{5}{c}{$^{98}$Zr} \\
        \hline
        \noalign{\smallskip}J$^\pi$ & E$_\gamma$ [keV] & $\tau_{\gamma\gamma}$ [ps] & $\tau_{\gamma}$ [ps] & $\tau_{lit.}$ [ps] \\
        \noalign{\smallskip} \hline    
        \noalign{\smallskip}$2^+_1$ & 1222.9(1) & 7.2(10) & & 3.79(79)~\cite{Singh2018}, 10(2)~\cite{Karayonchev2020}, $\leq$6.0~\cite{Ansari2017}, $\geq$0.68~\cite{Witt2018} \\
        $4^+_1$ & 620.5(2) & 5.51(94) & & 7.5(14)~\cite{Singh2018}, 13(5)~\cite{Karayonchev2020}, $\leq$15.0~\cite{Ansari2017}, 29(7)~\cite{Bettermann2010} \\
        $6^+_1$ & 647.58(3) & 3.16(57) & 2.82(31) & 2.63(89)~\cite{Singh2018}, $\leq$14~\cite{Bettermann2010}  \\
        $8^+_1$ & 725.4(1) & & 1.95(30) & 2.82(68)~\cite{Smith2012}  \\
        $10^+_1$ & 768.4(1) & & & 2.05(48)~\cite{Smith2012} \\

        \noalign{\smallskip}
        \hline \hline
        
        \noalign{\smallskip}\multicolumn{5}{c}{$^{100}$Zr} \\
        \hline
        \noalign{\smallskip}J$^\pi$ & E$_\gamma$ [keV] & $\tau_{\gamma\gamma}$ [ps] & $\tau_{\gamma}$ [ps] & $\tau_{lit.}$ [ps] \\
        \noalign{\smallskip} \hline
        \noalign{\smallskip} $2^+_1$  & 212.61(4) &  & & 1020(40)~\cite{Smith1996} 928(75)~\cite{Smith2002} 840(20)~\cite{Ansari2017}\\
        $4^+_1$ & 351.97(1) & 34.4(27) & 36.9(6) & 53(6)~\cite{Ohm1989} 53.4(5)~\cite{Smith2002} 37(4)~\cite{Ansari2017}  \\
        $6^+_1$ & 497.36(5) & 6.37(78) & 6.11(33) & 7.0(16)~\cite{Smith2002} 12(5)~\cite{Ansari2017}  \\
        $8^+_1$ & 625.55(5) & 1.66(40) & 1.32(19) & 2.55(30)~\cite{Smith1996,Smith2012} 2.49(25)~\cite{Smith2002} \\
        $10^+_1$ & 739.0(1) & & 0.72(15) & 1.08(12)~\cite{Smith1996,Smith2012} \\
        \noalign{\smallskip}
        \hline \hline
        
        \noalign{\smallskip}\multicolumn{5}{c}{$^{102}$Zr} \\
        \hline
        \noalign{\smallskip}J$^\pi$ & E$_\gamma$ [keV] & $\tau_{\gamma\gamma}$ [ps] & $\tau_{\gamma}$ [ps] & $\tau_{lit.}$ [ps] \\
        \noalign{\smallskip} \hline
        \noalign{\smallskip} $2^+_1$ & 151.8(1) & & & 2600(500)~\cite{Raman2001} 3610(430)~\cite{Browne2015_AP} 2914(87)~\cite{Ansari2017}\\
        $4^+_1$ & 326.5(2) & 41.6(39) & 45.9(13)  & 46.0(71)~\cite{Ansari2017} \\
        $6^+_1$ & 486.5(2) & 5.6(11) & 5.52(33) & $\leqslant$12~\cite{Ansari2017}  \\
        $8^+_1$ & 630.1(5) & 2.5(10) & 1.18(21) & 2.01(30)~\cite{Smith1996,Smith2012}  \\
        $10^+_1$ & 756.6(5) & & 1.27(52) & 0.77(12)~\cite{Smith1996,Smith2012} \\
        \noalign{\smallskip}
        \hline \hline
        
        \noalign{\smallskip}\multicolumn{5}{c}{$^{104}$Zr} \\
        \hline
        \noalign{\smallskip}J$^\pi$ & E$_\gamma$ [keV] & $\tau_{\gamma\gamma}$ [ps] & $\tau_{\gamma}$ [ps] & $\tau_{lit.}$ [ps] \\
        \noalign{\smallskip} \hline
        \noalign{\smallskip} $2^+_1$  & 139.3(3) &  &  &  2900(250)~\cite{Browne2015}\\
        $4^+_1$ & 312.2(3) & &  43.4(51) & \\
        $6^+_1$ &  473.7(3) & & 4.2(16) & \\
        $8^+_1$ & 624.4(3) & & & 1.91(29)~\cite{Smith1996,Smith2012} \\
        $10^+_1$ & 765.1(3) & & & 0.67(10)~\cite{Smith1996,Smith2012}  \\
        \noalign{\smallskip} \hline
\end{tabular}
\end{center}
\end{table*}

\section{Results}\label{results}

The lifetimes resulting from the present analysis are reported in \tablename~\ref{tab:results_lifetime} and compared with the previous experimental values.
The comparison reveals discrepancies between older RDDS measurements in $\gamma$-single mode and the present measurement utilizing $\gamma\gamma$ coincidences.
RDDS lifetime measurements in $\gamma$-single mode may overestimate the lifetime of an excited state if not all the feeding transitions can be properly taken into account. 
This difficulty is prevented by $\gamma\gamma$-coincidence measurements since the gate on a direct feeder permits the complete control of the population of the state. 
In general, the high efficiency and resolving power of the present experimental setup ensures a better control of the feeding resulting in a reasonable agreement between $\gamma\gamma$-coincidence and $\gamma$-single measurements.
However, when available, measurements in the $\gamma\gamma$-coincidence mode are always more reliable: at the price of a loss of statistical accuracy they are less affected by systematic uncertainties, which are often challenging to quantify in $\gamma$-single measurements.

From the measured lifetimes, transition probabilities can be calculated if the relevant branching and mixing ratios are known. 
The transitions of interest in this work are all of stretched E2 characters. 
The obtained values are corrected to take into account the internal conversion process and reported in \tablename~\ref{tab:results_BE2} both for the lifetimes measured in $\gamma\gamma$ coincidences and in $\gamma$-single mode. 
The resulting B(E2) values are compared with three recent theoretical predictions: the available values from the work of Ref.~\cite{Gavrielov2022} using the interacting boson model with configuration mixing (IBM-CM),  calculations from the Monte-Carlo Shell Model (MCSM) approach~\cite{Togashi2016,SabaThesis} and new results from symmetry-conserving configuration mixing (SCCM) calculations performed using the generator coordinate method framework
with Hartree-Fock-Bogoliubov states found with variation
after particle number projection (PN-VAP)~\cite{Rodriguez2010,Robledo2019}.
The experimental reduced transition probabilities for $^{98}$Zr, $^{100}$Zr, $^{102}$Zr and $^{104}$Zr are compared with the theoretical predictions (also given in \tablename~\ref{tab:results_BE2}) and shown in \figurename~\ref{fig:BE2_Zr}.

\section{Discussion}\label{discussion}

In the following our new experimental results are compared with theoretical predictions and differences with previous experimental results are discussed.

\subsection*{$^{98}$Zr}

Being near the critical point of the QPT in Zr isotopes, $^{98}$Zr presents a complicated level structure, as the result of the competition between different configurations present at low excitation energies. 
This makes the lifetime analysis in $\gamma$-single mode extremely complicated, especially at low spin. 
For this reason, the lifetime of the $2_1^+$ and the $4_1^+$ excited states has been extracted only with the $\gamma\gamma$-coincidence mode with a gate on a direct feeding transition.
Our results provide the unique possibility to reliably assign the lifetime of the $2_1^+$ and the $4_1^+$ excited states.
For the $2_1^+$ state we measured a value of 7.2(1.0)~ps by setting a gate on the $4_1^+ \rightarrow 2_1^+$ feeder. We find this value in between the two more recent results $\tau=3.8(8)$~ps~\cite{Singh2018} and $\tau=10(2)$~ps~\cite{Karayonchev2020}. 
The measurement from P.~Singh \textit{et al.}~\cite{Singh2018} was performed with the RDDS technique by using the same reaction and beam energy as in this work. 
However, a less efficient HPGe array enabled only $\gamma$-single measurements, which seems to overestimate the contribution of the observed transitions feeding the $2_1^+$ state, thus resulting in a shorter lifetime. 
The measurements from V.~Karayonchev \textit{et al.}~\cite{Karayonchev2020} were also performed with the RDDS technique, but the nuclear states of interest were populated via a two-neutron transfer reaction. 
This analysis also used the $\gamma$-single mode and the larger value of $\tau$ for the $2_1^+$ state suggests that in this case it was not possible to account for all feeding transitions of the state. However, this result~\cite{Karayonchev2020} agrees within the statistical error with the one in $\gamma\gamma$-coincidence mode from our experiment.
Without any influence from feeding corrections, the result from the present $\gamma\gamma$-coincidence analysis can be considered more reliable than previous results from $\gamma$-single measurements. 
The obtained B(E2) value of 1.5(2)~W.u. indicates both a single-particle nature for the $2_1^+ \to 0_1^+$ transition and a different deformation of the $2_1^+$ state with respect to a spherical ground-state, in line with the $2_1^+$ state being a member of a moderately deformed band based on the $0_2^+$ state, as already discussed in Ref.~\cite{Singh2018}. 
The MCSM (\textless~1~W.u.) and IBM-CM (1.35~W.u.) calculations correctly reproduce the small transition strength. 
The SCCM results show a triaxial-prolate (triaxial-oblate) $0^{+}_{1}$ ($2^{+}_{1}$) state that largely overestimate the experimental deformation. As a result, a more collective B(E2)-value is obtained (60 W.u.) for the $2_1^+ \rightarrow 0_1^+$ transition.

The lifetime of the $4_1^+$ state, $\tau\,$=~5.5(9)~ps, is compatible with the result from P.~Singh \textit{et al.}~\cite{Singh2018}, while the value from V.~Karayonchev \textit{et al.}~\cite{Karayonchev2020} seems again to be overestimated. 
Also the lifetime result $\tau\,$=~3.2(6)~ps of the $6_1^+$ state agrees with the one of Ref.~\cite{Singh2018}.
The presently obtained lifetimes correspond to a \hbox{B(E2;~$4_1^+ \to 2_1^+)$} value of 43(9)~W.u. and to a \hbox{B(E2;~$6_1^+ \to 4_1^+)$} value of 85(15)~W.u. which confirms the interpretation provided in Ref.~\cite{Singh2018} indicating that the $4_1^+$ state belongs to a well-deformed structure possibly based on the $0_3^+$ state. 
The large \hbox{B(E2;~$4_1^+ \to 2_1^+)$} value indicates that this well-deformed structure based on the $0_3^+$ state presents a signiﬁcant mixing with the moderately deformed conﬁguration based on the $0^+_2$ state, as pointed out in Ref.~\cite{Singh2018,Wu2004}.
This increase is in agreement with the predictions from all the three IBM-CM, MCSM and SCCM calculations (see top panel in figure \ref{fig:BE2_Zr}). 
For the latter method, a transition from a triaxial-prolate $2^{+}_{1}$ state to axial-prolate $4^{+}_{1}-10^{+}_{1}$ states is found. These states belong to a first excited band with a larger deformation than the ground state and which becomes yrast at J$\,^\pi =4^{+}$. As mentioned above, the deformation of all states in $^{98}$Zr are overestimated by the present SCCM calculations, also those of the axial-prolate deformed band.

Higher-lying states were populated with less statistics and a lifetime measurement for the $8_1^+$ state was only possible in $\gamma$-single mode. Such short lifetimes of 1-2~ps are at the limit of the sensitivity of the RDDS method, because the measurement can be perturbed by effects from the slowing-down process of the recoiling nuclei in the degrader. 
Our result agrees within 2 standard deviations with the measurement of Ref.~\cite{Smith2012}, which uses the Doppler Shift Attenuation Method (DSAM)~\cite{Nolan1979}, the most suitable technique to measure lifetimes down to hundreds of fs. 
The resulting B(E2)=78(12)~W.u., as well as the B(E2)=55(13)~W.u. measured in the same DSAM work~\cite{Smith2012} for the $10_1^+ \rightarrow 8_1^+$ transition, are well described by the constant trend of the IBM-CM predictions.
The results corroborate the interpretation of coexistence between a deformed configuration that becomes yrast at spin $4^+$ with a spherical configuration dominating the ground state.

\begin{table*}[htbp]
\begin{center}
\begin{threeparttable}
    \caption{\label{tab:results_BE2}Reduced transition probabilities, B(E2;~$J^\pi \to (J-2)^\pi$), calculated from the lifetimes of excited states in $^{98}$Zr, $^{100}$Zr, $^{102}$Zr and $^{104}$Zr determined in the present work from $\gamma\gamma$-coincidence and $\gamma$-singles data. The transition energies are taken from Ref.~\cite{nndc}. }
    \vspace{1mm}
\begin{tabular}{p{.8cm}p{2cm}p{2cm}p{2.cm}p{1.9cm}p{1.9cm}p{1.9cm}}
  \hline \hline
  \noalign{\smallskip}\multicolumn{1}{c}{$^{98}$Zr} &\multicolumn{6}{c}{B(E2$\downarrow$) [e$^2$b$^2$]\tnote{[1]}} \\
  \hline
  \noalign{\smallskip}J$^\pi$ &  E$_\gamma$ [keV] & $\gamma\gamma$ coinc. & $\gamma$ single & IBM-CM & SCCM & MCSM \\
  \noalign{\smallskip} \hline    
  \noalign{\smallskip}$2^+_1$\tnote{[a]}  &1222.9(1)  & 0.0041(6) & & 0.0036 & 0.1628 & 0.0018 \\
  $4^+_1$\tnote{[b]} & 620.5(2) & 0.145(25) & & 0.1825 & 0.3007 & 0.2760 \\
  $6^+_1$ & 647.58(3) & 0.227(41) & 0.254(28) & 0.2064 & 0.5225 & 0.2762 \\
  $8^+_1$ & 725.4(1) & & 0.209(32) & 0.1928 & 0.7003 & \\
  $10^+_1$ & 768.4(1) & & & 0.1519 & 0.7838 & \\

  \noalign{\smallskip}
  \hline \hline  
  \noalign{\smallskip}\multicolumn{1}{c}{$^{100}$Zr} &\multicolumn{6}{c}{B(E2$\downarrow$) [e$^2$b$^2$]\tnote{[2]}} \\
  \hline
  \noalign{\smallskip}J$^\pi$ &  E$_\gamma$ [keV] & $\gamma\gamma$ coinc. & $\gamma$ single & IBM-CM & SCCM & MCSM \\
  \noalign{\smallskip} \hline
  \noalign{\smallskip} $2^+_1$ & 212.61(4) & & & 0.1985 & 0.2604 & 0.2480 \\
  $4^+_1$ & 351.97(1) & 0.434(34) & 0.405(7) & 0.3336 & 0.4144 & 0.3529 \\
  $6^+_1$ & 497.36(5) & 0.420(51) & 0.438(24) & 0.3557 & 0.4927 & 0.3192 \\
  $8^+_1$ & 625.55(5) & 0.510(120) & 0.645(93) & 0.3391 & 0.5464 & \\
  $10^+_1$ & 739.0(1) & & 0.510(110) & 0.2923 & 0.5873 & \\
  \noalign{\smallskip}
  \hline \hline
        
  \noalign{\smallskip}\multicolumn{1}{c}{$^{102}$Zr} &\multicolumn{6}{c}{B(E2$\downarrow$) [e$^2$b$^2$]\tnote{[3]}} \\
  \hline
  \noalign{\smallskip}J$^\pi$ &  E$_\gamma$ [keV]& $\gamma\gamma$ coinc. & $\gamma$ single & IBM-CM & SCCM & MCSM \\
  \noalign{\smallskip} \hline
  \noalign{\smallskip} $2^+_1$ & 151.8 & & & 0.3624 & 0.2856 & 0.2935 \\
  $4^+_1$  & 326.5(2) & 0.521(49) & 0.473(13) & 0.5017 & 0.4235 & 0.4191 \\
  $6^+_1$ & 486.5(2) & 0.530 (110) & 0.541(32) & 0.5209 & 0.4819 & 0.4425 \\
  $8^+_1$ & 630.1(5) & 0.330(130) & 0.700(120) & 0.4983 & 0.5177 & \\
  $10^+_1$ & 756.6(5) & & 0.260(110) & 0.4445 & 0.5415 & \\
  \noalign{\smallskip}
  \hline \hline

  \noalign{\smallskip}\multicolumn{1}{c}{$^{104}$Zr} &\multicolumn{6}{c}{B(E2$\downarrow$) [e$^2$b$^2$]\tnote{[4]}} \\
  \hline
  \noalign{\smallskip}J$^\pi$ & E$_\gamma$ [keV] & $\gamma\gamma$ coinc. & $\gamma$ single & IBM-CM & SCCM & MCSM \\
  \noalign{\smallskip} \hline
  \noalign{\smallskip} $2^+_1$ & 139.3 & & & 0.4184 & 0.378 & \\
  $4^+_1$  & 312.2(3) & & 0.623(73) & 0.5840 & 0.5588 & \\
  $6^+_1$ &  473.7(3) & & 0.810(310) & 0.6130 & 0.6377 & \\
  $8^+_1$ &  624.4(3) & & & 0.5956 & 0.6908 & \\
  $10^+_1$ & 765.1(3) & & & 0.5491 & 0.7329 & \\
  \noalign{\smallskip} \hline
\end{tabular}
\begin{tablenotes}
      \begin{small} 
      \item [a] From the measured lifetime $\tau_{\gamma\gamma}(2_1^+)\,$=~7.2(10) ps the resulting B(E2) value for the $2_1^+ \to 0_2^+$ transition at 368.8(1)~keV is 0.0428~e$^2$b$^2$.
      \item [b] From the measured lifetime $\tau_{\gamma\gamma}(4_1^+)\,$=~5.51(94) ps the resulting B(E2) value for the $4_1^+ \to 2_2^+$ transition at 252.7(2)~keV is 0.2546~e$^2$b$^2$.
      \item [1] The reported B(E2) values ($x$) in e$^2$b$^2$ for A=98 correspond to $x\cdot 10^4$/26.84 W.u.
      \item [2] The reported B(E2) values ($x$) in e$^2$b$^2$ for A=100 correspond to $x \cdot 10^4$/27.57 W.u.
      \item [3] The reported B(E2) values ($x$) in e$^2$b$^2$ for A=102 correspond to $x\cdot10^4$/28.31 W.u. 
      \item [4] The reported B(E2) values ($x$) in e$^2$b$^2$ for A=104 correspond to $x \cdot10^4$/29.05 W.u.
      \end{small}
    \end{tablenotes}
\end{threeparttable}
\end{center}
\end{table*}

\subsection*{$^{100}$Zr, $^{102}$Zr, $^{104}$Zr}

\begin{figure}[htbp]
\centerline{%
\includegraphics[width=0.51\textwidth]{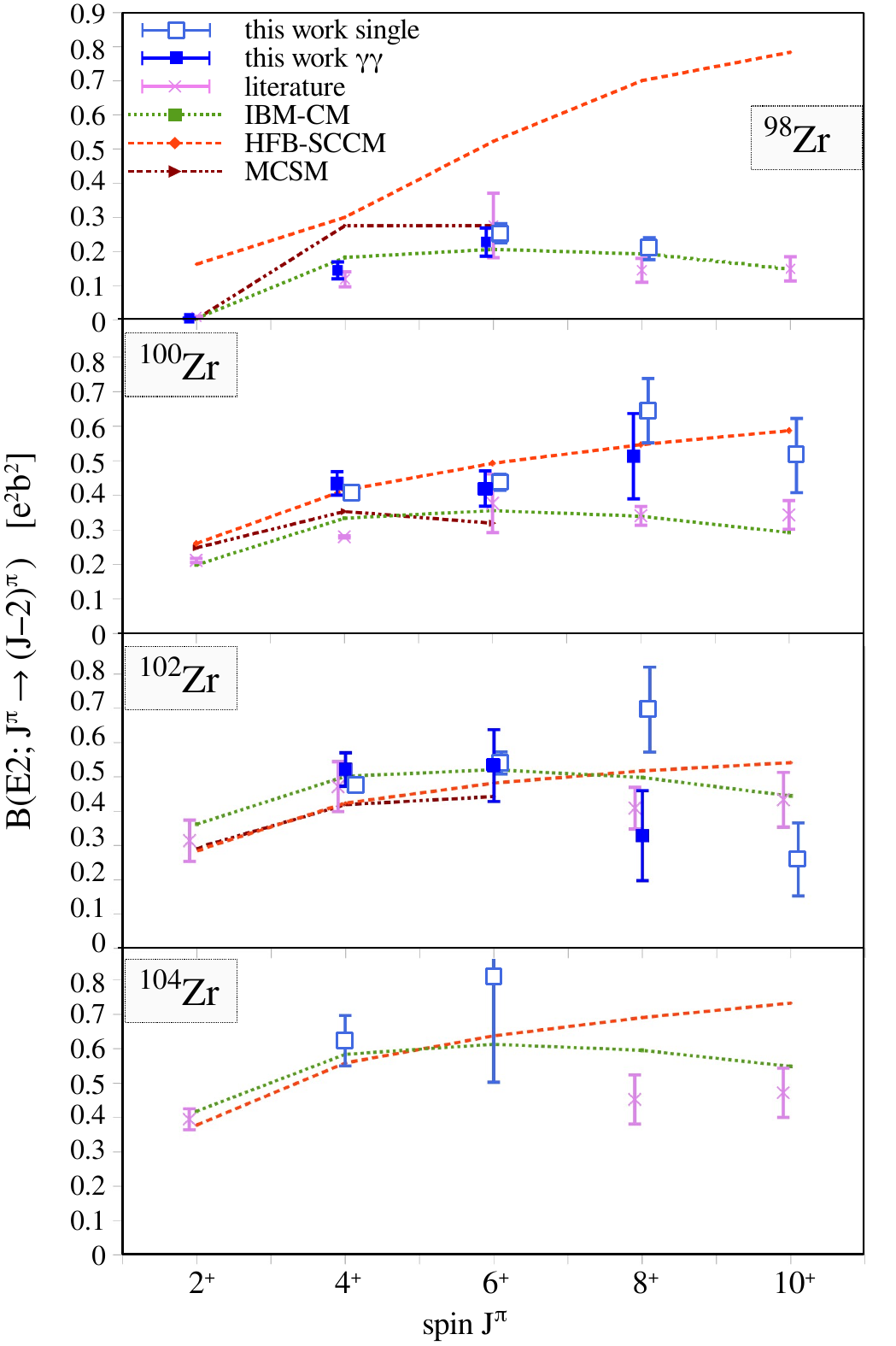}}
\caption{Transition probabilities measured from this experiment, as in \tablename~\ref{tab:results_BE2}, for $^{98}$Zr, $^{100}$Zr, $^{102}$Zr, $^{104}$Zr compared with the calculations of the IBM-CM~\cite{Gavrielov2022}, SCCM~\cite{Robledo2019} and MCSM~\cite{Togashi2016}. Literature values from the Evaluated Nuclear Data Files~\cite{nndc} and Ref.~\cite{Ansari2017} are also reported for comparison. }
\label{fig:BE2_Zr}
\end{figure}

For $^{100}$Zr, $^{102}$Zr and $^{104}$Zr the lifetime of the $2^+_1$ state could not be measured with the RDDS technique in this experiment since it is too long for the chosen plunger distances (the B(E2;~$2_1^+ \rightarrow 0_1^+$) values in $^{100,102,104}$Zr results presented in \figurename~\ref{fig:BE2_Zr} are taken from the literature). 

The lifetimes for the $4^+_1$ and the $6^+_1$ states in $^{100}$Zr and $^{102}$Zr were measured both with the $\gamma$-single and $\gamma\gamma$-coincidence mode and there is an agreement with one standard deviation between the values. 
However, the lifetimes of the $4^+$ states measured with $\gamma\gamma$ coincidences are systematically shorter than the values measured in $\gamma$-single mode, from both the present and previous measurements. 
Gamma-single measurements can be in fact affected by unobserved feeding and this effect is enhanced for the lowest-lying excited states.
The lifetime of the $4^+_1$ state in $^{100}$Zr is a good example: the contribution of the $4_2^+ \rightarrow 4_1^+$ feeding transition, visible with a small intensity in the $\gamma$-ray spectra, cannot be taken into account in the analysis since the shifted component of this transition at 850~keV overlaps with the stopped component of the $(12_1)^+ \rightarrow 10_1^+$ transition at 841~keV. 
The lifetime of the $4^+_1$ state in $^{100}$Zr was first measured in 1989 by H.~Ohm and collaborators~\cite{Ohm1989} with a fast-timing technique in $\beta\gamma$ coincidences following the decay of an isomeric state of $^{100}$Y. 
This result, $\tau = 53(6)$~ps, has been later confirmed by a differential plunger measurement in the work of Smith~\textit{et al.} in 2002~\cite{Smith2002}, obtaining $\tau = 53.4(5)$~ps. 
A recent fast-timing experiment with LaBr$_3$(Ce) scintillators measured a shorter lifetime for this state, $\tau = 37(4)$~ps~\cite{Ansari2017}. 
This measurement was carried out by gating on the feeding $\gamma$-ray transition, ensuring that the result is not affected by the feeding contribution.
Our measurement, $\tau = 34(2)$~ps, is in agreement with the latter result~\cite{Ansari2017} and calls for a reconsideration of the previously accepted lifetime of $\tau = 53.4(5)$~ps for this state, which value is probably influenced by the feeding from higher-lying states.
The lifetime of the $6^+_1$ excited state in $^{100}$Zr has been measured by Smith~\textit{et al.}~\cite{Smith2002}, $\tau = 7(2)$~ps, and Ansari~\textit{et al.}~\cite{Ansari2017}, $\tau = 12(5)$~ps, the former being in perfect agreement with our result in $\gamma\gamma$ coincidences, $\tau = 6.4(8)$~ps. 

In $^{102}$Zr only the measurement of $\tau=46(7)$~ps of the $4_1^+$ state and an upper limit of 12~ps for the $6_1^+$ state are present in the literature from the work of Ansari \textit{et al.}~\cite{Ansari2017}. Our results are in agreement with these values and set a lifetime of  $\tau=5.6(11)$~ps (in $\gamma\gamma$ coincidences) for the $6_1^+$ state for the first time.

The lifetimes of the $8^+_1$ and $10^+_1$ excited states in $^{100}$Zr and $^{102}$Zr are in agreement with previous measurements obtained with the Doppler Shift Attenuation Method (DSAM) in the work of Smith~\textit{et al.} in 2012~\cite{Smith2012}.
The RDDS technique manifests its limitations for lifetimes of the order of 1-2~ps: this can explain the difference in the $\gamma$-single and $\gamma\gamma$-coincidence measurements for the $8_1^+$ state (even though these are compatible considering the uncertainties). 

For $^{104}$Zr, lifetimes of $\tau=43(5)$~ps for the $4^+_1$ and of $\tau=4(2)$~ps for the $6^+_1$ state are measured in $\gamma$-single mode for the first time in this experiment. \\

For $^{100}$Zr, $^{102}$Zr and $^{104}$Zr the B(E2$\downarrow$) values show an increase of collectivity between the 2$^+$ and the 4$^+$ state and a smaller variation at higher spins.
The increase of the average B(E2) values from $^{100}$Zr to $^{104}$Zr indicates that the collectivity and is also increasing as neutrons are added beyond N=60, which supports the downward trend of the $2^+_1$ excitation energies (\figurename~\ref{fig:energy_BE2}). 
The results from the SCCM method show a rotational trend which is in better agreement with the experimental data for $^{100}$Zr, $^{102}$Zr and $^{104}$Zr than in the $^{98}$Zr case. 
These calculations also predict shape coexistence and shape mixing, although the ground-state bands are all well-deformed (mostly triaxial-prolate) and do not cross with other bands. 
Hence, \figurename~\ref{fig:BE2_Zr} shows that the B(E2) values continuously increase with the angular momentum. 
This is not the case for the IBM-CM results nor for the experimental values found in the literature, where a decrease in B(E2) values are observed in the $8^{+}$ and $10^{+}$ transitions. The present RDDS data in $^{100}$Zr suggests the opposite while no definitive conclusions can be drawn from $^{102,104}$Zr. 
Therefore, future B(E2) measurements for $8^{+}$ and $10^{+}$ transitions would help to disentangle whether these nuclei keep their rotational character at those angular momenta or some structural change appears to modify slightly such a rotational trend.


\section{Conclusion} \label{conclusion}

This work presents the measurement of lifetimes of 15 excited states in $^{98}$Zr, $^{100}$Zr, $^{102}$Zr and $^{104}$Zr by using the Recoil Distance Doppler Shift technique. The previously unknown lifetime of 3 excited states in $^{102}$Zr and $^{104}$Zr were measured for the first time, while 9 lifetimes of excited states were measured for the first time in $\gamma\gamma$ coincidences, avoiding the influence from feeding transitions. 
The use of the $\gamma\gamma$-coincidence technique has been particularly advantageous in the case of $^{98}$Zr, where the feeding pattern is intricate. 
From the lifetime results, transition probabilities are extracted and compared with different state-of-the-art theoretical calculations. 
The results are in agreement with a large increase of the transition probabilities between low-lying excited states starting from $^{100}$Zr, indicating a change in shape with respect to the lighter Zr isotopes. 
The calculated transition probabilities show an increase of deformation between the $2^+$ and the $4^+$ state and a rotational behaviour at higher spins.
In particular, for $^{98}$Zr the significant change of the B(E2) values between the $2^+_1$ and the $4^+_1$ state confirms the predicted coexistence of different shapes for this system.\\

The authors would like to thank the AGATA and VAMOS++ collaborations. We are grateful for the help from J. Goupil and the GANIL technical staff for their work in setting up the apparatuses and the good quality beam. 
The authors would also like to thank G. Fremont for preparing the target and degrader foils. 
G.P. and S.A. acknowledge M. Siciliano for the fruitful discussions and the help with lifetime analysis.
This work has been partly funded (G.P.) by the P2IO LabEx (ANR-10-LABX-0038) in the framework Investissements d’Avenir (ANR-11-IDEX- 0003-01) managed by the French Agence Nationale de la Recherche (ANR). 
The work of T.R.R. is supported by the Spanish MICINN under PRE2019-088036. 
T.R.R. gratefully thanks the support from the GSI-Darmstadt computing facility. 
A.G, J.S.H., V.M and L.G.P acknowledge the support of Norwegian Research Council, projects 240104, 263030, and 325714. 
Z.P gratefully thanks the support of STFC (UK). 
The work of P.-A.S. is supported by BMBF under grant NuSTAR.DA 05P15RDFN1, contract PN 23.21.01.06 sponsored by the Romanian Ministry of Research, Innovation and Digitalization. 
A.E, L.G., J.J., L.K. and J.-M.R. acknowledge the BMBF Verbundprojekt 05P2021 (ErUM-FSP T07) grant No. 05P21PKFN1. 
S.L. acknowledges funding from the European Union-NextGenerationEU, through the National Recovery and Resilience Plan of the Republic of Bulgaria, project No. BG-RRP-2.004-0008-C01. 
The work of A.M.B. and E.R.G. was financially supported by the Science and Technology Facility Council (STFC) Grant No. ST/L005840/1. 
M.S. has been supported by the OASIS project no. ANR-17-CE31-0026 and by he U.S. Department of Energy, Office of Science, Office of Nuclear Physics, under contract number DE-AC02-06CH11357. \\



\end{document}